\documentclass{elsart}

\usepackage{graphicx}
\usepackage{amssymb}

 \newcommand{\dtot}[2] { \frac{d {#1} } {d {#2}} }
 \newcommand{\dtoth}[2] { {d {#1} }/{d {#2}} }
 
 \newcommand{\dpar}[2] { \frac{\partial {#1} } {\partial {#2}} }
 \newcommand{\ndpar}[3] { \frac{\partial^{#3} {#1} } {\partial #2 ^{#3} }}
 
 \newcommand{\dparh}[2] { {\partial {#1} }/{\partial {#2}} }
 
 \newcommand{\OL}[1] {\textrm{\emph{\large O}}\!\left({#1}\right)}

 \newcommand{\II}[0] {\mbox{i}}

 \newcommand{\dis}[0] {\displaystyle}
 
 \newcommand{\eqref}[1] {(\ref{#1})}

\begin{document}

\begin{frontmatter}

 \title{Self-similar Radiation from Numerical Rosenau-Hyman Compactons}

\author[RUS]{Francisco Rus},
 \ead{rusman@lcc.uma.es}
\author[VILLA]{Francisco R. Villatoro\corauthref{COR}}
 \ead{villa@lcc.uma.es}

\address[RUS]{E.T.S. Ingenier\'{\i}a Inform\'atica,
 Dept. Lenguajes y Ciencias de la Computaci\'on, \\
Universidad de M\'alaga, Campus de Teatinos, 29071 M\'alaga,
Spain}

 \corauth[COR]{Corresponding author:
 Tel.: +34-95-2132096; fax: +34-95-2132816.}
\address[VILLA]{E.T.S. Ingenieros Industriales,
 Dept. Lenguajes y Ciencias de la Computaci\'on, \\
Universidad de M\'alaga, Campus de El Ejido, 29013 M\'alaga,
Spain}

\begin{abstract}
The numerical simulation of compactons, solitary waves with compact
support, is characterized by the presence of spurious phenomena, as
numerically-induced radiation, which is illustrated here using four
numerical methods applied to the Rosenau-Hyman $K(p,p)$ equation.
Both forward and backward radiations are emitted from the compacton
presenting a self-similar shape which has been illustrated
graphically by the proper scaling. A grid refinement study shows
that the amplitude of the radiations decreases as the grid size
does, confirming its numerical origin. The front velocity and the
amplitude of both radiations have been studied as a function of both
the compacton and the numerical parameters. The amplitude of the
radiations decreases exponentially in time, being characterized by a
nearly constant scaling exponent. An ansatz for both the backward
and forward radiations corresponding to a self-similar function
characterized by the scaling exponent is suggested by the present
numerical results.
\end{abstract}

\begin{keyword}
Compactons \sep Numerical radiation \sep Self-similarity  \sep
Rosenau-Hyman Equation

%\MSC 35Q51, 81T80

\end{keyword}

\end{frontmatter}

\section{Introduction}
\label{Intro}

Compactons are travelling wave solutions with compact support
resulting from the balance of both nonlinearity and nonlinear
dispersion. Compacton solutions have been first found in a
generalized Korteweg-de Vries equation with nonlinear dispersion,
the so-called (focusing) $K(p,p)$ compacton equation of Rosenau and
Hyman~\cite{RosenauHyman1993}, given by
\begin{equation}
 \label{RH}
 \dpar{u}{t} - c_0\,\dpar{u}{x} + \dpar{u^p}{x} + \ndpar{u^p}{x}{3} = 0,
 \label{eq:Knn}
\end{equation}
where $u(x,t)$ is the wave amplitude, $x$ is the spatial coordinate,
$t$ is time, and $c_0$ is a constant velocity used here in order to
stop the compacton when required. Compactons are classical solutions
of this equation only for $1<p\le 3$, otherwise they are
``non-classical" or weak solutions. In fact, compacton solutions of
Eq.~\eqref{RH}, for $p\not\in\{-1,0,1\}$, can be written
as~\cite{Rosenau1998}
\begin{equation}
 \label{compactonsolution}
 u_c(x,t) =
  \left\{
  \begin{array}{lr}
  \alpha^\mu\,
  \cos^{2\mu} \left(
        \beta\,\xi(x,t)
       \right), \qquad \qquad
  &
  \dis |\xi(x,t)|\le \frac{\pi}{2\,\beta},
  \\
  0,
  &
 \mbox{otherwise},
 \end{array}
 \right.
\end{equation}
\[
 \xi(x,t) = x-x_0-(c-c_0)\,t,
 \quad
 \alpha=\frac{2\,c\,p}{p+1},
 \quad
 \beta=\frac{p-1}{2\,p},
 \quad
 \mu = \frac{1}{(p-1)},
\]
where $c$ is the compacton velocity and $x_0$ the position of its
maximum at $t=0$. Note that the compacton has $k$ continuous
derivatives at its both edges when $p=(2+k)/k$.

Compactons have multiple applications in Physics. Rosenau-Hyman (RH)
equation~\eqref{RH} was discovered as a simplified model to study
the role of nonlinear dispersion on pattern formation in liquid
drops~\cite{RosenauHyman1993}, being also proposed in the analysis
of patterns on liquid surfaces~\cite{Ludu1998}. Equations with
compacton solutions have also found applications such as the
lubrication approximation for thin viscous
films~\cite{Bertozzi1996}, semiclassical models for Bose-Einstein
condensates~\cite{Kovalev1998}, long nonlinear surface waves in a
rotating ocean when the high-frequency dispersion is
null~\cite{Grimshaw1998}, the pulse propagation in ventricle-aorta
system~\cite{Kardashov2006}, dispersive models for magma
dynamics~\cite{Simpson2007}, or, even, particle wavefunctions in
nonlinear quantum mechanics~\cite{Caparelli1998}. RH equation is
also the continuous limit of the discrete equations of a nonlinear
lattice~\cite{RosenauHyman1993}. In nonlinear lattices the
propagation of compacton-like kinks has been observed using
mechanical~\cite{Dusuel1998}, electrical~\cite{Comte2002,Comte2006},
and magnetic~\cite{Prilepsky2006} analogs. Recently, RH equation has
been generalized using a cosine nonlinearity in order to model the
dispersive coupling in chains of oscillators resulting in the
so-called phase compactons and kovatons~\cite{Pikovsky2005}, having
application in superconducting Josephson junction transmission
lines~\cite{Pikovsky2006}. Finally, let us remark that the general
$K(p,q)$ equation, with $p\ne q$, may also show elliptic function
compactons~\cite{RosenauHyman1993,Rosenau2000,Cooper2006}, and that
recent interest is focusing on multidimensional
compactons~\cite{Rosenau2006,RosenauHyman2007}.

The numerical simulation of the propagation of nonlinear waves
presents several numerically induced phenomena, such as spurious
radiation, artificial dissipation, and errors in group velocity. The
numerical analysis of compactons are not free of these spurious
phenomena. In fact, the numerical solution of compacton equations is
a very challenging problem presenting several numerical difficulties
which has not been currently
explained~\cite{Rosenau2000,deFrutosSanzSerna1995,SaucezEtAl2004}.

The numerical simulation of compactons by means of pseudospectral
methods in space require the addition of artificial dissipation
(hyperviscosity) using high-pass
filters~\cite{RosenauHyman1993,RosenauHyman2007,ChertockLevy2001,CooperHyman2001}
in order to obtain stable results without appreciable spurious
radiation. In fact, using those methods,
Ref.~\cite{RosenauHyman1993} shows that compactons collide
elastically, without visible radiation. However, after the
collision, compactons show a phase shift and a small-amplitude,
zero-mass, compact ripple is generated, which slowly decomposes into
tiny compacton-anticompacton pairs. Numerical simulations without
high-pass filtering show that these compact ripples present internal
shock layers~\cite{deFrutosSanzSerna1995,RusVillatoro2005}. The main
drawback of current (filtered) pseudospectral methods is the
inability to show high-frequency phenomena. Particle methods based
on the dispersive-velocity method have been proposed to cope with
these features~\cite{ChertockLevy2001}, but their preservation of
the positivity of the solutions is another clear disadvantage, since
after compacton collisions the solution may change sign.

Both finite element and finite difference methods without
high-frequency filtering have also been proposed. In finite element
methods both a Petrov-Galerkin method using the product
approximation developed by Sanz-Serna and
Christie~\cite{deFrutosSanzSerna1995}, and a standard method based
on piecewise polynomials discontinuous at the element
interfaces~\cite{LevyShuYan2004} have been used. Second-order finite
difference methods~\cite{IsmailTaha1998,HanXu2007}, high-order
Pad\'e methods~\cite{RusVillatoro2005}, and the method of lines with
adaptive mesh refinement~\cite{SaucezEtAl2004} has also been applied
with success. These methods also require artificial dissipation to
simulate the generation of shocks after compacton interactions,
which is usually incorporated by a linear fourth-order derivative
term. Such term introduces a plateau tail whose amplitude has been
calculated for the $K(2,2)$ equation by Pikovsky and
Rosenau~\cite{Pikovsky2006} by means of a variational perturbation
theory for compactons.

The main drawback in the numerical simulation of compacton
propagation without high-frequency filtering is the appearance of
spurious radiation, even in one-compacton solutions, as first shown
by the authors in Ref.~\cite{RusVillatoro2006} by means of using the
fourth-order Petrov-Galerkin finite element method developed by de
Frutos, L\'opez-Marcos and Sanz-Serna~\cite{deFrutosSanzSerna1995}.
Both backward and forward propagating wavepackets of radiation are
emitted from the compacton, having a very small amplitude, in fact,
more than six orders of magnitude smaller than the compacton
amplitude in current simulations.

The main goal of this paper is a detailed analysis of the numerical
origin and the main properties of the radiation emitted by
compactons observed in Ref.\cite{RusVillatoro2006}. First, in order
to illustrate the universality of this phenomena, three additional
numerical methods are considered: the second order finite difference
method developed by Ismail and Taha~\cite{IsmailTaha1998} and two
Pad\'e methods of sixth and eighth order developed by Rus and
Villatoro~\cite{RusVillatoro2005}. Second, the numerical origin of
the radiation is clarified by means of a grid refinement study.
Third, in order to check if the origin of the radiation is due to
the jump at the edge of the compacton suffered by its second-order
derivative, the $K(p,p)$ equations having compactons showing jumps
in their first- to eight-order derivatives, i.e., with $p\in\{ 3, 2,
5/3, 3/2, 7/5, 4/3, 9/7, 5/4\}$, are also considered. And fourth,
the graphical illustration of the self-similarity of the radiation
is complemented with the numerical determination of their front
velocity, wavepacket mean amplitude, and self-similar scaling
exponents.

The contents of this paper are as follows. Next Section presents the
four numerical methods for the Rosenau-Hyman $K(p,p)$ equation
analyzed in this paper. Section~\ref{Results} presents our results
on the properties characterizing both the forward and the backward
numerically-induced radiation wavepackets generated during the
propagation of one-compacton solutions. Finally, the last section is
devoted to some conclusions.

\section{Numerical methods}
\label{NumericalMethods}

Let us consider the numerical solution of the RH Eq.~\eqref{RH} by
means of the method of lines in time and several Pad\'e
approximations in space. Periodic boundary conditions in the
interval $x\in[0,L]$ are used as an approximation of the initial
value problem in the whole real line. Let us take the fixed grid
spacing $\Delta x=L/M$, the nodes $x_m=m\,\Delta x$, for $m=0, 1,
\ldots, M$, and a general Pad\'e method written as
\begin{equation}
 \label{Pade}
 {\mathcal{A}_i(\mbox{E})}\,\dtot{U_{m}}{t}
         - c_0\,{\mathcal{B}_i(\mbox{E})} \, (U_m)
         + {\mathcal{B}_i(\mbox{E})} \, (U_m)^p
         + {\mathcal{C}_i(\mbox{E})} \,(U_m)^p
          = 0,
\end{equation}
where $U_m\approx u(x_m)$, $\mbox{E}$ is the shift operator, i.e.,
$\mbox{E}\,U_m=U_{m+1}$, the first and second derivatives are
rationally approximated by means of
$\mathcal{B}_i(\mbox{E})/\mathcal{A}_i(\mbox{E})$ and
$\mathcal{C}_i(\mbox{E})/\mathcal{A}_i(\mbox{E})$, respectively,
and $i$ indicates the method among those studied in this paper.

\textbf{Method 1}. The finite difference method developed by Ismail
and Taha~\cite{IsmailTaha1998} is given by
\[
 {\mathcal{A}_1(\mbox{E})}
 =
  \mathcal{I},
\]
\[
 {\mathcal{B}_1(\mbox{E})}
 =
 \frac{ -\mbox{E}^{-1}+\mbox{E}^{1}
      }{2\,\Delta x},
\]
\[
 {\mathcal{C}_1(\mbox{E})}
 =
 \frac{ -\mbox{E}^{-2}+2\,\mbox{E}^{-1}-2\,\mbox{E}^{1}+\mbox{E}^{2}
      }{2\,\Delta x^3},
\]
where $\mathcal{I}$ is the identity operator. In this case,
method~\eqref{Pade} is second-order accurate in space since
\[
 \frac{\mathcal{B}_1(\mbox{E})}{\mathcal{A}_1(\mbox{E})}\, u =
 \dpar{u}{x}+ \frac{\Delta x^2}{6}\,\ndpar{u}{x}{3}+\OL{\Delta x^4},
\]
and
\[ \frac{\mathcal{C}_1(\mbox{E})}{\mathcal{A}_1(\mbox{E})}\, u =
 \ndpar{u}{x}{3} +  \frac{\Delta x^2}{4}\,\ndpar{u}{x}{5}+\OL{\Delta
 x^6}.
\]

\textbf{Method 2}. The finite element method developed by de Frutos
et al.~\cite{deFrutosSanzSerna1995} is obtained by using
\[
 {\mathcal{A}_2(\mbox{E})}
 =
  \frac{ \mbox{E}^{-2}+26\,\mbox{E}^{-1}+66+26\,\mbox{E}^{1}+\mbox{E}^{2}
       }{120},
\]
\[
 {\mathcal{B}_2(\mbox{E})}
 =
 \frac{ -\mbox{E}^{-2}-10\,\mbox{E}^{-1}+10\,\mbox{E}^{1}+\mbox{E}^{2}
      }{24\,\Delta x},
\]
and ${\mathcal{C}_2(\mbox{E})} = {\mathcal{C}_1(\mbox{E})}$, where
${\mathcal{B}_2(\mbox{E})}/{\mathcal{A}_2(\mbox{E})}$ and
${\mathcal{C}_2(\mbox{E})}/{\mathcal{A}_2(\mbox{E})}$ are sixth- and
fourth-order approximations to, respectively, the first- and
third-order derivatives in Eq.~\eqref{RH}, in fact
\[
 \frac{\mathcal{B}_2(\mbox{E})}{\mathcal{A}_2(\mbox{E})}\, u =
 \dpar{u}{x}+ \frac{\Delta x^6}{5040}\,\ndpar{u}{x}{7}+\OL{\Delta x^8},
\]
and
\[ \frac{\mathcal{C}_2(\mbox{E})}{\mathcal{A}_2(\mbox{E})}\, u =
 \ndpar{u}{x}{3} -  \frac{\Delta x^4}{240}\,\ndpar{u}{x}{7}+\OL{\Delta
 x^6}.
\]
Hence this method is fourth-order accurate in space.

\textbf{Method 3}. A Pad\'e method introduced in
Ref.~\cite{RusVillatoro2005} which approximates the third- and
first-order derivatives with, respectively, sixth- and fourth-order
of accuracy, given by
\[
 {\mathcal{A}_3(\mbox{E})}
 =
  \frac{ \mbox{E}^{-2}+56\,\mbox{E}^{-1}+126+56\,\mbox{E}^{1}+\mbox{E}^{2}
       }{240},
\]
${\mathcal{B}_3(\mbox{E})} = {\mathcal{B}_2(\mbox{E})}$, and
${\mathcal{C}_3(\mbox{E})} = {\mathcal{C}_1(\mbox{E})}$. In fact,
Taylor series expansion yields
\[
 \frac{\mathcal{B}_3(\mbox{E})}{\mathcal{A}_3(\mbox{E})}\, u =
 \dpar{u}{x}+ \frac{\Delta x^4}{240}\,\ndpar{u}{x}{5}+\OL{\Delta x^6},
\]
and
\[ \frac{\mathcal{C}_3(\mbox{E})}{\mathcal{A}_3(\mbox{E})}\, u =
 \ndpar{u}{x}{3} -  \frac{\Delta x^6}{60480}\,\ndpar{u}{x}{9}+\OL{\Delta
 x^8}.
\]

\textbf{Method 4}. Another Pad\'e method also introduced in
Ref.~\cite{RusVillatoro2005} with an eighth-order accurate
approximation to the first derivative in Eq.~\eqref{RH}, obtained
by means of
\[
 {\mathcal{A}_4(\mbox{E})}
 =
  \frac{ \mbox{E}^{-2}+16\,\mbox{E}^{-1}+36+16\,\mbox{E}^{1}+\mbox{E}^{2}
       }{70},
\]
\[
 {\mathcal{B}_4(\mbox{E})}
 =
 \frac{ -5\,\mbox{E}^{-2}-32\,\mbox{E}^{-1}+32\,\mbox{E}^{1}+5\,\mbox{E}^{2}
      }{84\,\Delta x},
\]
and ${\mathcal{C}_4(\mbox{E})} = {\mathcal{C}_1(\mbox{E})}$. This
method is only of second-order for the third-order derivative, as
shown by Taylor series expansion. Concretely,
\[
 \frac{\mathcal{B}_4(\mbox{E})}{\mathcal{A}_4(\mbox{E})}\, u =
 \dpar{u}{x}- \frac{\Delta x^8}{44100}\,\ndpar{u}{x}{9}+\OL{\Delta x^{10}},
\]

and \[ \frac{\mathcal{C}_4(\mbox{E})}{\mathcal{A}_4(\mbox{E})}\, u =
 \ndpar{u}{x}{3} -  \frac{\Delta x^2}{28}\,\ndpar{u}{x}{5}+\OL{\Delta
 x^4}.
\]

For sufficiently regular solutions of Eq.~\eqref{RH}, Methods~2
and~3 are fourth-order accurate, and Methods~1 and~4 only of
second-order. Here on, Methods~1--4 are referred to as Ismail, de
Frutos, Pad\'e-6, and Pad\'e-8, respectively. Note that Methods~1--4
may be classified in function of the numerical order of
approximation for the first and third derivatives in its local
truncation error terms as $(2,2)$, $(6,4)$, $(4,6)$, and $(8,2)$,
respectively.

In this paper, the integration in time of Equation~\eqref{Pade} is
obtained by means of both the trapezoidal rule,
\begin{eqnarray}
 \label{Trapezoidal}
 &&
 {\mathcal{A}_i(\mbox{E})}\,\frac{U_{m}^{n+1} - U_{m}^{n} }{\Delta t}
         -c_0 \,\mathcal{B}_i(\mbox{E})\,
         \frac{ U_{m}^{n+1}+ U_{m}^{n} }{2}
 \nonumber
 \\ &&
 \phantom{ {\mathcal{A}_i(\mbox{E})}\,\frac{U_{m}^{n+1} - U_{m}^{n} }{\Delta t} }
         + \left(
           {\mathcal{B}_i(\mbox{E})}
         + {\mathcal{C}_i(\mbox{E})}
           \right)\,\frac{ \left(U_{m}^{n+1}\right)^p + \left(U_{m}^{n}\right)^p }{2}
          = 0,
\end{eqnarray}
and the implicit midpoint rule,
\begin{eqnarray}
 \label{Midpoint}
 &&
 {\mathcal{A}_i(\mbox{E})}\,\frac{U_{m}^{n+1} - U_{m}^{n} }{\Delta t}
        -c_0 \,\mathcal{B}_i(\mbox{E})\,
         \frac{ U_{m}^{n+1}+ U_{m}^{n} }{2}
 \nonumber
 \\ &&
 \phantom{ {\mathcal{A}_i(\mbox{E})}\,\frac{U_{m}^{n+1} - U_{m}^{n} }{\Delta t} }
          + \left(
           {\mathcal{B}_i(\mbox{E})}
         + {\mathcal{C}_i(\mbox{E})}
           \right)\, \left( \frac{ U_{m}^{n+1} + U_{m}^{n} }{2}\right)^p
          = 0,
\end{eqnarray}
where $t^n=n\,\Delta t$ and $U_m^n\approx u(x_m,t^n)$. Both methods
are second-order accurate in time and yields implicit equations
solved by using the Newton's method.

The linear stability analysis by the von Neumann method for the
methods developed in this section applied to the linearization of
Eq.~\eqref{RH} shows its unconditional (linear)
stability~\cite{deFrutosSanzSerna1995,RusVillatoro2005,IsmailTaha1998}.
Note that the usefulness of this linear stability analysis may be
criticized when applied to a highly nonlinear problem as
Eq.~\eqref{RH}, however, it is standard in a numerical analysis
context. In fact, the solution of the four methods may blow-up for
some $\Delta x$ and $\Delta t$ due to nonlinear instabilities whose
analysis is outside the scope of this paper.

Equation~\eqref{RH} has four invariants $I_j=\int \phi_j(u)\,dx$,
where $\phi_1=u$, $\phi_2=u^{p+1}$, $\phi_3=u\,\cos(x)$, and
$\phi_4=u\,\sin(x)$. Methods~1--4 preserve exactly the first
invariant of the $K(p,p)$ equation, however, the other three
invariants are not exactly preserved, but instead only well
preserved~\cite{deFrutosSanzSerna1995,RusVillatoro2005,IsmailTaha1998}.

\section{Presentation of results}
\label{Results}

Extensive numerical simulations of the $K(p,p)$ equation with
several $p$ using either the trapezoidal or the implicit midpoint
rule yield practically the same results for all of Methods~1--4, at
least for $\Delta t > \Delta x/10$, hence, only results using the
implicit midpoint rule are hereafter presented and discussed. For
the sake of brevity, unless anything else is stated, the following
figures and tables only show the results for the $K(2,2)$
equation.\footnote{Supplementary material with figures and tables
presenting results for the $K(p,p)$ equation with $p\in\{ 3, 2, 5/3,
3/2, 7/5, 4/3, 9/7, 5/4\}$ may be found in the web page
\texttt{http://www.lcc.uma.es/$\sim$rusman/invest/compact/Compactons.htm}.}

\begin{figure}
 \begin{center}
  \includegraphics[width=11cm]{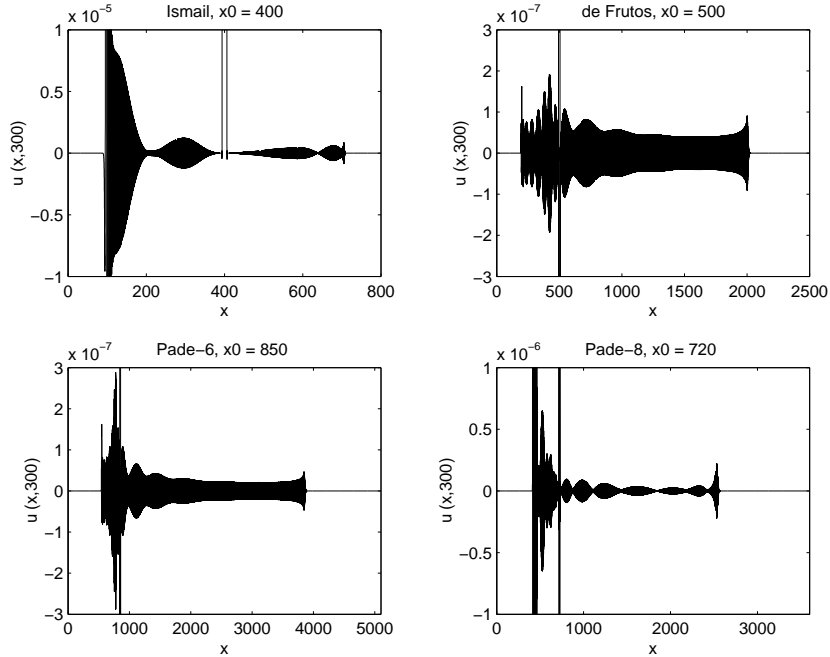}
 \end{center}
  \caption{Zoom in of snapshots  at $t=300$ of the radiation
  generated at
  both sides of a compacton  of the $K(2,2)$ equation
  propagating with $c=1$
  numerically calculated using $c_0=1$, $\Delta x=0.05$, and $\Delta t=0.1$
  by means of
  Ismail (top left plot),
  de Frutos (top right plot),
  Pad\'e-6 (bottom left plot),
  and~Pad\'e-8 (bottom right plot) methods,
  initially located at,  respectively, $x_0=400$, 500,
  850, and~720.}
  \label{fig:BothRadiations}
\end{figure}

The four plots in Fig.~\ref{fig:BothRadiations} show vertical zooms
of the solution of the $K(2,2)$ equation at $t=300$ for an initial
condition given by one compacton with velocity $c=1$ initially
located at $x_0=400$, 500, 850, and~720 for, respectively, Ismail
(top left plot), de Frutos (top right plot), Pad\'e-6 (bottom left
plot), and~Pad\'e-8 (bottom right plot) methods. The four plots in
Fig.~\ref{fig:BothRadiations} clearly show that two wavepackets of
radiation are generated from the compacton, here referred to as
forward and backward radiation corresponding to that propagating to
the right and to the left, respectively, of the compacton. Note that
the initial position of the compactons is not the same in all the
plots in order to avoid that the backward (forward) radiation cross
the left (right) boundary reappearing through the other one due to
the periodic boundary conditions used in the simulations. Note also
the use of $c_0=c$ in order to stop the compacton and highlight the
relative velocity of both wavepackets of radiation generated during
its propagation.

The plots in Fig.~\ref{fig:BothRadiations} show that the amplitude
of both wavepackets is very small compared with that of the
compacton, being that of the backward radiation two orders of
magnitude larger than that of the forward one for Ismail
(Fig.~\ref{fig:BothRadiations}, top left plot) and Pad\'e-8 (bottom
right plot), but only several times largest for the other two
methods. The backward radiation has a steeper front than that of the
forward one for all the methods and a front velocity smaller (in
absolute value) than the forward one for de Frutos (top right plot),
Pad\'e-6 (bottom left plot), and~Pad\'e-8 (bottom right plot)
methods, being approximately equal for Ismail (top left plot) one.
Note that, for long time integrations under periodic boundary
conditions, both radiation wavepackets collide resulting in a
background dominated by the backward radiation, whereon the
compacton propagates, due to its robustness, without appreciable
change on its parameters.

\begin{figure}
 \begin{center}
  \includegraphics[height=9cm]{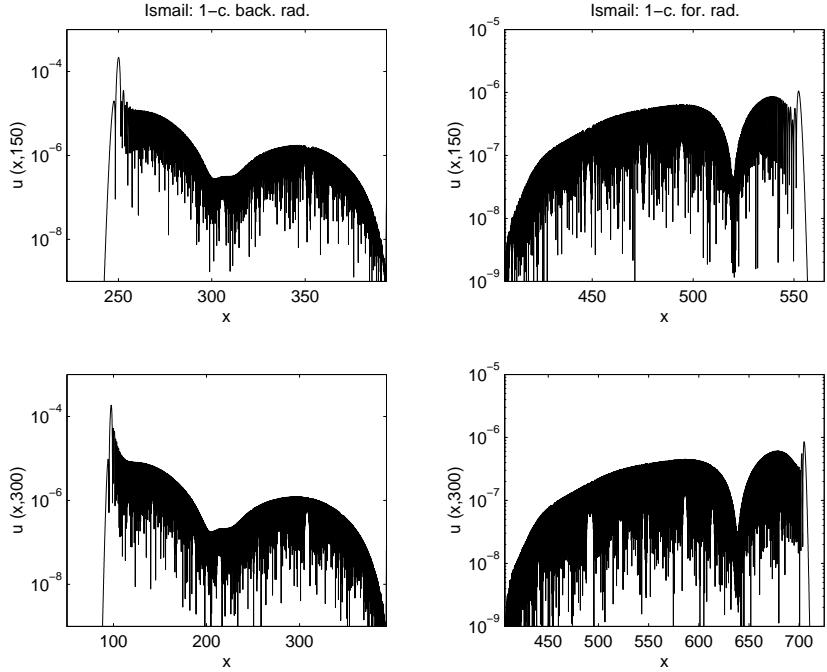} %% estiradas4x4Ismail
 \end{center}
  \caption{Backward (left plots) and forward (right ones)
  radiation generated by  a compacton
   of the $K(2,2)$ equation numerically propagating with Ismail method with $\Delta x=0.05$, $\Delta t=0.1$,
  and $c_0=c=1$ at two instants of time, $t=150$ (top plots)
  and $t=300$ (bottom ones), highlighting their self-similarity.}
 \label{Ismail:similar}
\end{figure}

\begin{figure}
 \begin{center}
  \includegraphics[height=9cm]{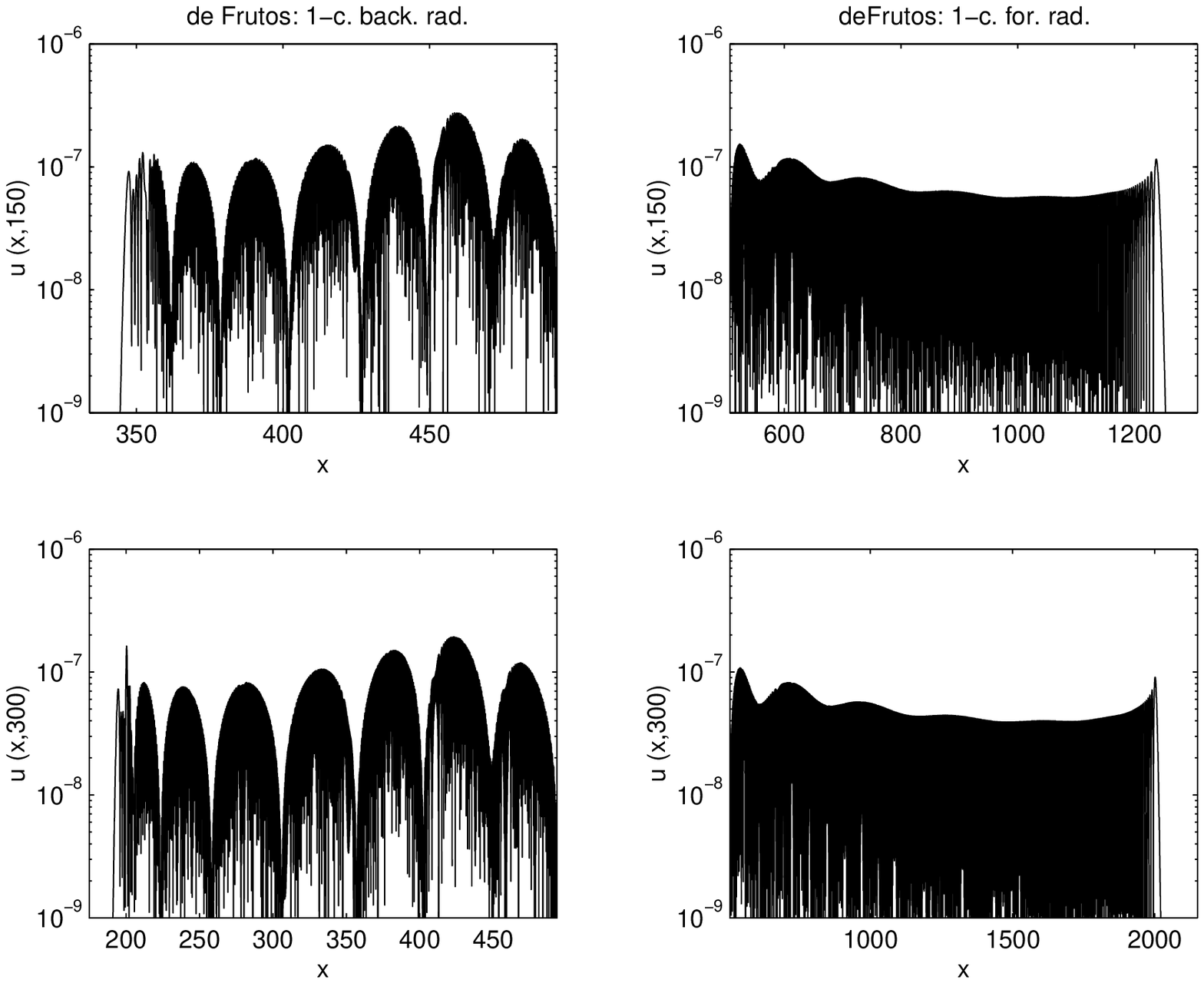} %% estiradas4x4deFrutos
 \end{center}
  \caption{Backward (left plots) and forward (right ones)
  radiation generated by  a compacton
   of the $K(2,2)$ equation numerically propagating with de Frutos method with $\Delta x=0.05$, $\Delta t=0.1$,
  and $c_0=c=1$ at two instants of time, $t=150$ (top plots)
  and $t=300$ (bottom  ones), highlighting their self-similarity.}
 \label{deFrutos:similar}
\end{figure}

\begin{figure}
 \begin{center}
  \includegraphics[height=9cm]{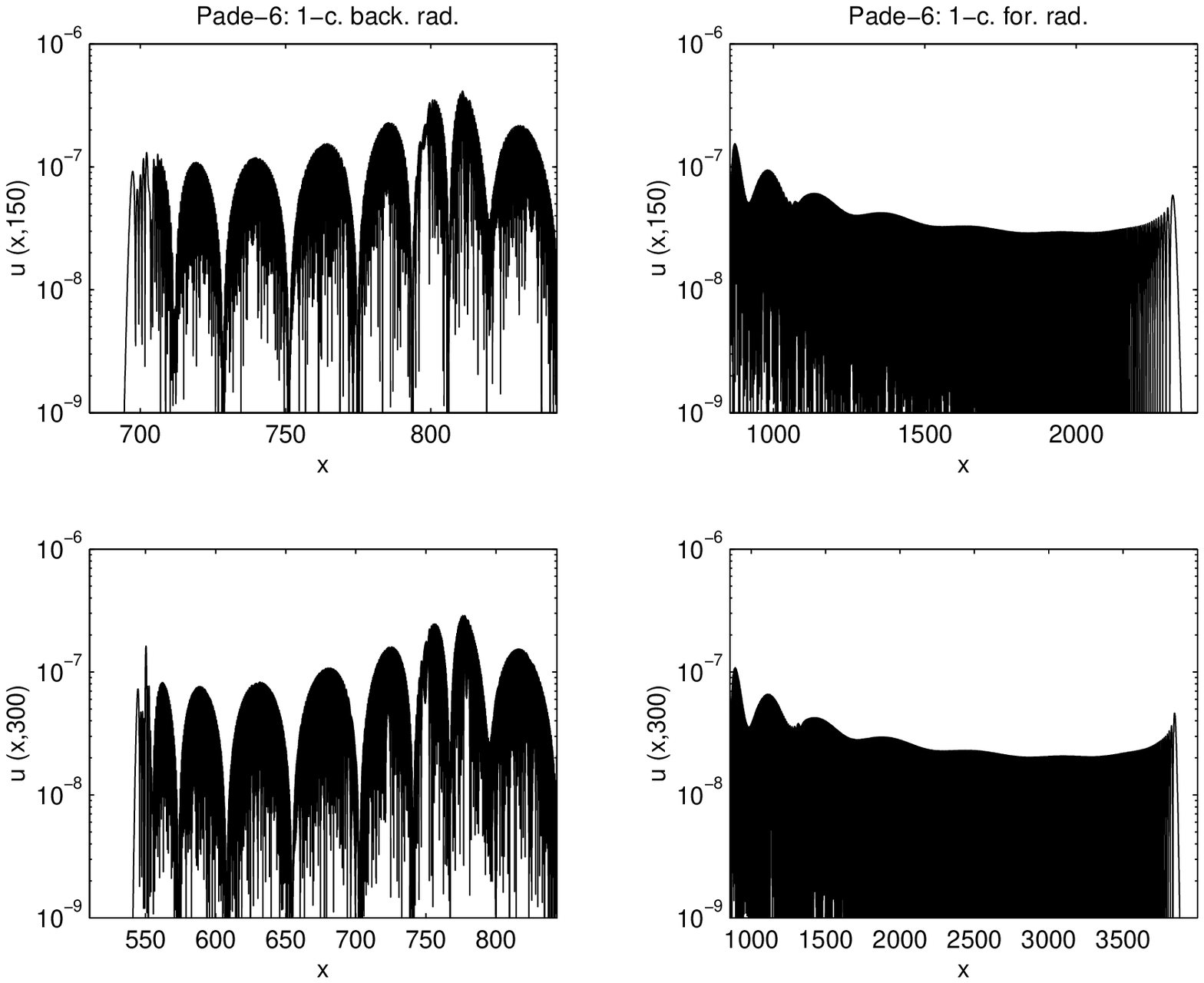} %% estiradas4x4BIS
 \end{center}
  \caption{Backward (left plots) and forward (right ones)
  radiation generated by  a compacton
   of the $K(2,2)$ equation numerically propagating with Pad\'e-6 method with $\Delta x=0.05$, $\Delta t=0.1$,
  and $c_0=c=1$ at two instants of time, $t=150$ (top plots)
  and $t=300$ (bottom ones), highlighting their self-similarity.}
 \label{Pade6:similar}
\end{figure}

\begin{figure}
 \begin{center}
  \includegraphics[height=9cm]{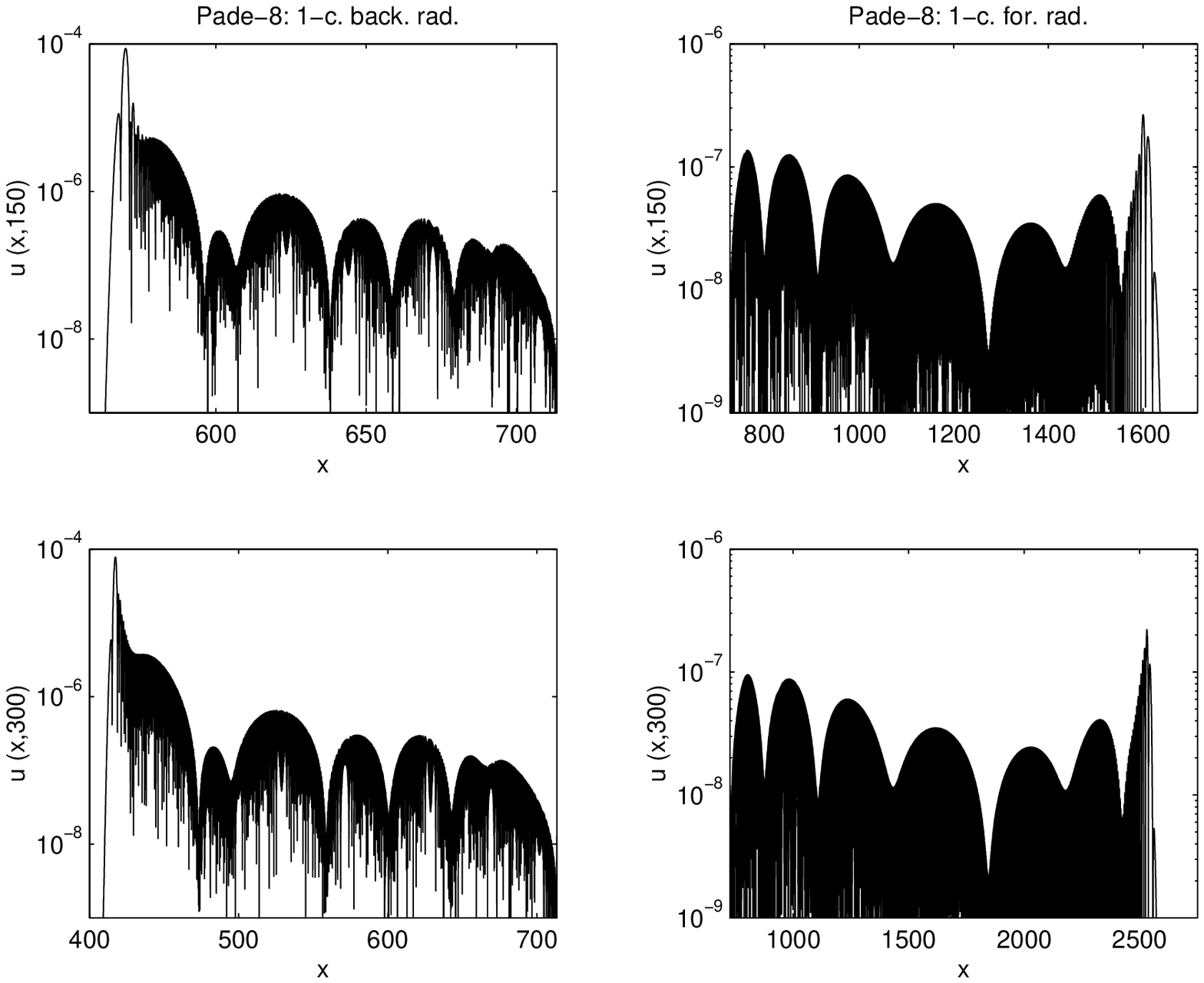} %% estiradas4x4TRIS
 \end{center}
  \caption{Backward (left plots) and forward (right ones)
  radiation generated by  a compacton
   of the $K(2,2)$ equation numerically propagating
   with Pad\'e-8 method with $\Delta x=0.05$, $\Delta t=0.1$,
  and $c_0=c=1$ at two instants of time, $t=150$ (top plots)
  and $t=300$ (bottom ones), highlighting their self-similarity.}
 \label{Pade8:similar}
\end{figure}

The more interesting and noticeable property of both backward and
forward compacton radiation is their self-similarity.
Figures~\ref{Ismail:similar}, \ref{deFrutos:similar},
\ref{Pade6:similar}, and~\ref{Pade8:similar} show the absolute value
of both the forward (right plots) and the backward (left ones)
radiation for, respectively, Ismail, de Frutos, Pad\'e-6, and
Pad\'e-8 methods at time $t=150$ (top plots) and $t=300$ (bottom
ones). In the plots of
Figs.~\ref{Ismail:similar}--\ref{Pade8:similar}, the horizontal axis
is selected in order to best illustrate the self-similarity of the
wavepacket envelope of both numerically-induced radiations by
graphical comparison of the top and bottom plots. As shown in
Figs.~\ref{Ismail:similar}--\ref{Pade8:similar} the envelope shape
of both the forward and the backward radiations is highly dependent
on the method and, not illustrated in the plots, on its parameters
$\Delta x$ and  $\Delta t$, and the compacton velocity $c$.

A possible origin of the self-similar radiation may be the jump
experienced by the second-order derivative of the $K(2,2)$ compacton
at their edges. However, extensive numerical simulations using
Methods~1--4 show that the self-similarity of the radiation is also
present in the propagation of $K(p,p)$ compactons with $k$
continuous derivatives at its both edges, i.e., for $p=(2+k)/k$. The
only cases in which the self-similarity is not clearly visible are
for $p \gtrapprox 1$, for which the amplitude of the radiation is
comparable with the tolerance used in the iterations of the Newton
method. In such cases, the radiation near the compacton is degraded
by noise, apparently introduced by round-off errors, destroying the
self-similarity and, for long-time integrations, blowing up the
solution.

The numerical origin of the spurious radiation observed in the
simulations is illustrated in Tables~\ref{AmplitudesDX}
and~\ref{AmplitudesDT}, which show the amplitude of both the
backward ($u_b$) and forward ($u_f$) radiation for the four methods
studied in this paper as a function of $\Delta x$ and $\Delta t$,
respectively. This amplitude has been determined by finding the
first local maximum of the wavepacket starting from the front of the
wavepacket using a five point rule, i.e., three nodes where the
function increases followed by two nodes where it decreases, being
the amplitude value that of the central node.
Tables~\ref{AmplitudesDX} and~\ref{AmplitudesDT} clearly show that
the amplitude of both radiations decreases with decreasing $\Delta
x$ but remains practically constant with $\Delta t$. The numerical
origin of the radiations appear to be the numerical approximation of
the spatial derivatives. The last column of Table~\ref{AmplitudesDX}
shows the exponent $q$ such that the amplitude of the radiations are
$\OL{\Delta x^q}$, calculated by means of linear regression. This
exponent may clarified whether the approximation of either the first
or the third derivatives in Eq.~\eqref{eq:Knn} is the only
responsible of any of these radiations. However, the results shown
in Table~\ref{AmplitudesDX} are not conclusive in this respect and
the radiations appear to be the result of the trade-off between the
local truncation error of both derivatives. Similar results have
been obtained for the other $K(p,p)$ equations studied in this
paper.

\begin{table}
 \caption{Amplitude at $t=150$ of both the backward ($u_b$) and forward
 ($u_f$) radiation for a compacton of the $K(2,2)$ equation
 with velocity $c=1$ as a function of $\Delta x$ using $\Delta t=0.05$,
 $c_0=c$, and $x\in[0,2500]$. The asterisks indicate solutions which
 blow up. Linear regression is used to obtain $q$ such that $u_f$ and $u_b$ are $\OL{\Delta x^q}$.
 }
{\small
\[
\begin{array}{|c|c|c|c|c|c|c|c|c|c|}
  \hline  %
  Meth. & \Delta x             & 0.2         & 0.1         & 0.05        & 0.025       & 0.0125      & q \\ \hline
  1 & u_f & 6.54\times 10^{-5} & 7.14\times 10^{-6} & 1.26\times 10^{-6} & 2.24\times 10^{-7} &   *          & 2.7 \\
    & u_b & 2.78\times 10^{-3} & 6.22\times 10^{-3} & 2.94\times 10^{-4} & 5.93\times 10^{-5} &    *         & 2.1 \\ \hline
  2 & u_f & 5.68\times 10^{-6} & 7.09\times 10^{-7} & 1.61\times 10^{-7} & 2.71\times 10^{-8} & 8.04\times 10^{-9} & 2.4 \\
    & u_b & 1.45\times 10^{-5} & 2.54\times 10^{-6} & 2.60\times 10^{-7} & 4.80\times 10^{-8} & 1.50\times 10^{-8} & 2.6 \\ \hline
  3 & u_f & 8.32\times 10^{-6} & 4.00\times 10^{-7} & 8.95\times 10^{-8} & 2.11\times 10^{-8} & 8.12\times 10^{-9} & 2.4 \\
    & u_b & 1.84\times 10^{-5} & 4.46\times 10^{-6} & 3.81\times 10^{-7} & 7.43\times 10^{-8} & 2.35\times 10^{-8} & 2.5  \\ \hline
  4 & u_f & 1.76\times 10^{-5} & 1.64\times 10^{-6} & 2.79\times 10^{-7} & 4.84\times 10^{-8} & 7.23\times 10^{-9} & 2.8  \\
    & u_b & 4.24\times 10^{-3} & 3.17\times 10^{-4} & 8.90\times 10^{-5} & 2.34\times 10^{-5} & 5.97\times 10^{-6} & 2.3  \\ \hline
 \end{array}
\]
}
 \label{AmplitudesDX}
\end{table}

\begin{table}
 \caption{Amplitude at $t=150$ of both the backward ($u_b$) and forward
 ($u_f$) radiation for a compacton of the $K(2,2)$ equation
 with velocity $c=1$ as a function of $\Delta t$ using $c_0=c$, $\Delta x=0.05$,
 and $x\in[0,2500]$. Linear regression is used to obtain $q$ such that $u_f$ and $u_b$ are $\OL{\Delta t^q}$.
 }
{\small
\[
\begin{array}{|c|c|c|c|c|c|c|c|c|}
  \hline  %
  Meth.& \Delta t             & 0.1         & 0.05        & 0.025       & 0.0125      & 0.00625 & q \\ \hline
  1 & u_f & 1.04\times 10^{-6} & 1.26\times 10^{-6} & 1.37\times 10^{-6} & 1.40\times 10^{-6} & 1.41\times 10^{-6} & -.10 \\
    & u_b & 2.17\times 10^{-4} & 2.94\times 10^{-4} & 1.53\times 10^{-3} & 1.44\times 10^{-3} & 1.21\times 10^{-3} & -.73 \\ \hline
  2 & u_f & 1.52\times 10^{-7} & 1.61\times 10^{-7} & 1.96\times 10^{-7} & 2.11\times 10^{-7} & 2.16\times 10^{-7} & -.14 \\
    & u_b & 2.72\times 10^{-7} & 2.60\times 10^{-7} & 4.18\times 10^{-7} & 7.88\times 10^{-7} & 7.70\times 10^{-7} & -.46 \\ \hline
  3 & u_f & 1.53\times 10^{-7} & 8.95\times 10^{-8} & 1.11\times 10^{-7} & 1.24\times 10^{-7} & 1.29\times 10^{-7} & .002 \\
    & u_b & 4.11\times 10^{-7} & 3.81\times 10^{-7} & 1.11\times 10^{-6} & 8.71\times 10^{-7} & 7.00\times 10^{-7} & -.27 \\ \hline
  4 & u_f & 2.65\times 10^{-7} & 2.79\times 10^{-7} & 2.75\times 10^{-7} & 2.81\times 10^{-7} & 2.93\times 10^{-7} & -.03 \\
    & u_b & 8.64\times 10^{-5} & 8.90\times 10^{-5} & 8.18\times 10^{-5} & 8.17\times 10^{-5} & 8.08\times 10^{-5} & .030 \\ \hline
 \end{array}
\]
}
 \label{AmplitudesDT}
\end{table}

%% Curvas de posicion del frente de onda => Velocidad parece constante

\begin{figure}
 \begin{center}
  \includegraphics[width=11.0cm]{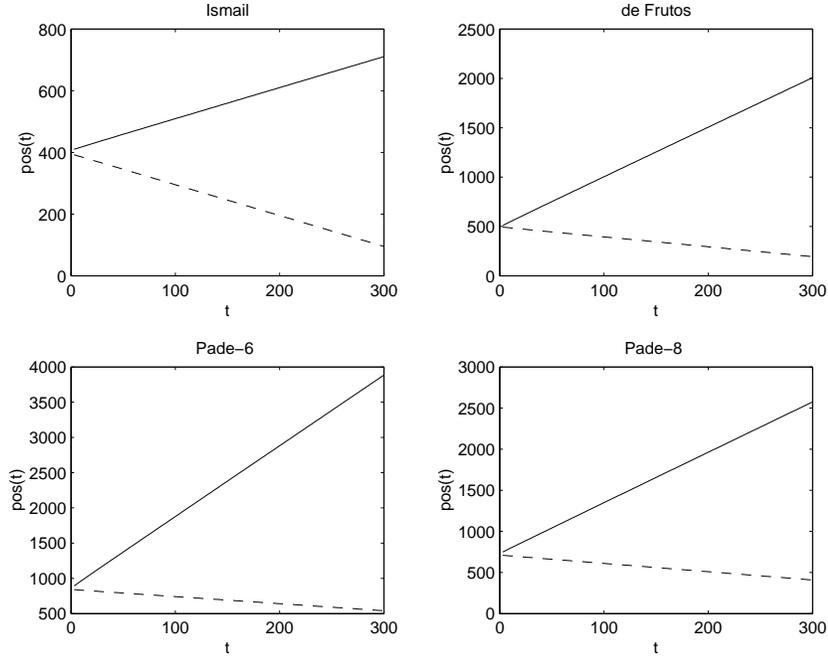} %% PosFrentes
 \end{center}
  \caption{Front velocity evolution
  of both forward (continuous line) and backward (dashed line)
  radiations for a $K(2,2)$ compacton numerically propagating as a function of time for Ismail (top left plot),
  de Frutos (top right), Pad\'e-6 (bottom left) and Pad\'e-8
 (bottom right) methods with $\Delta x=0.05$, $\Delta t=0.1$
 and $c_0=c=1$.}
  \label{fig:VelocitiesRadiation}
\end{figure}

The position of the left (right) front of the backward (forward)
radiation wavepackets for the Ismail (top left plot), de Frutos (top
right plot), Pad\'e-6 (bottom left plot) and Pad\'e-8 (bottom right
plot) methods for the $K(2,2)$ equation is shown in
Fig.~\ref{fig:VelocitiesRadiation}. This position has been
determined, using linear interpolation, as the ``first" point from
the outside of the wavepacket, i.e., from left to right (right to
left) for backward (forward) radiation, where the amplitude of the
solution is equal to an amplitude threshold, the half of the maximum
amplitude of the radiation at $t=300$. The four plots in
Fig.~\ref{fig:VelocitiesRadiation} clearly show a linear evolution
of the position of the front for both forward (continuous line) and
backward (dashed line) radiations. The velocity of the front, i.e.,
the slope of these curves, is nearly constant during propagation
being negative (positive) for backward (forward) radiation. The
constancy of the front velocities has also been observed in the
simulations of the $K(p,p)$ equation.

The front velocity of both the forward ($c_f$) and the backward
($c_b$) wavepackets relative to the velocity of the compacton may be
calculated by linear regression from the evolution in time of their
positions. Our extensive numerical experiments show that both the
forward ($c_f$) and backward ($c_b$) front velocities are linear
functions of the parameter $c_0$, being practically independent of
the parameters $\Delta x$ and $\Delta t$, and, also nearly
independent of $c$ and $p$, for a $K(p,p)$ compacton, with a small
percentage increase as $p$ decreases approaching unity.

Table~\ref{VelocidadesFrentesEnCo} shows both the forward ($c_f$)
and backward ($c_b$) front velocities for $c_0=c/2$, $c$, and $2c$,
for two values of $\Delta x$ and two values of $\Delta t$. The front
velocities depend linearly on $c_0$ instead on $c$, in fact,
$c_f\approx c_0$, $5\,c_0$, $10\,c_0$, and~$6\,c_0$ for
Methods~1--4, respectively, and $c_b\approx -c_0$ for the four
methods. This approximations are better as $\Delta x$ decreases. Let
us highlight that both front velocities are relative to that of the
compacton, therefore, in a rest frame of reference, where the
compacton propagates with its own velocity instead of being stopped
by the condition $c_0=c$, the backward radiation is generated in the
left edge of the compacton at $t=0$ and stretches as it propagates,
like a wake left in the track of the compacton during its
propagation signaling its initial position in the numerical
simulation.

\begin{table}
 \caption{Velocity of the front of both the backward and forward
 radiation for a $K(2,2)$ compacton with velocity $c=1$ as function of $c_0$
 in numerical simulations with two $\Delta x$, two $\Delta t$,
 $x\in[0,2500]$ and $t\in[0,100]$.
 }
{\small
\[
\begin{array}{|c|c|c|c|c||c|c|c||c|c|c|}
  \hline  %
  \multicolumn{2}{|c|}{}&\multicolumn{6}{c||}{ \Delta x=0.1}&\multicolumn{3}{c|}{ \Delta x=0.5} \\
  \cline{3-11}
  \multicolumn{2}{|c|}{}&\multicolumn{3}{c||}{\Delta t=0.025} & \multicolumn{6}{c|}{\Delta t=0.05}\\
  \hline
  Meth. & c_0 & 1/2     & 1    & 2    & 1/2     & 1    & 2    & 1/2     & 1    & 2    \\ \hline
  1 & c_f & * & 1.02 & 2.03 & * & 1.02 & 2.03 & * & 1.01 & 2.02 \\
    & c_b & * &-1.02 &-2.03 & * &-1.03 &-2.04 & * &-1.02 &-2.03  \\ \hline
  2 & c_f & 2.55    & 5.06 & 10.1 & 2.56    & 5.07 & 10.1 & 2.53    & 5.05 & * \\
    & c_b & -0.500  &-1.01 &-2.01 &-0.504   &-1.01 &-2.02 & -0.505  &-1.01 & *  \\ \hline
  3 & c_f & 5.09    & 10.1 & 20.1 & 5.09    & 10.1 & 20.2 & 5.06    & 10.1 & * \\
    & c_b & -0.672  &-1.25 &-2.05 &-0.634   &-1.03 &-2.02 & -0.514  &-1.01 & *  \\ \hline
  4 & c_f & 3.12    & 6.18 & 12.3 & 3.12    & 6.19 & 12.4 & 3.10    & 6.16 & * \\
    & c_b & -0.510  &-1.00 &-2.01 &-0.506   &-1.01 &-2.02 & -0.505  &-1.01 & *  \\ \hline
 \end{array}
\]
}
 \label{VelocidadesFrentesEnCo}
\end{table}

%% effect of variations of c_0

\begin{figure}
 \begin{center}
  \includegraphics[width=7.0cm]{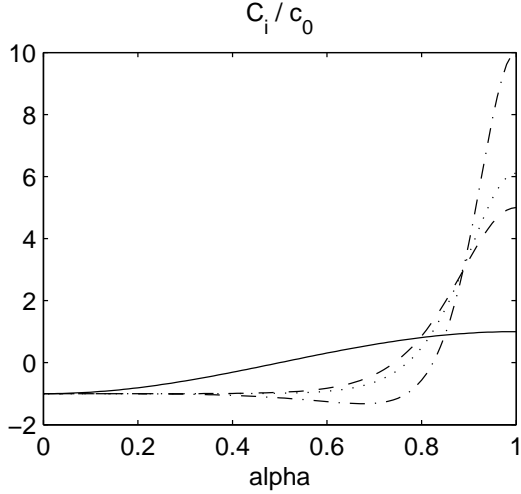} %% figureCg
 \end{center}
  \caption{Group velocity of Method~1~(solid line),
  2~(dashed line), 3~(dash-dotted line), and~4~(dotted line)
  for the linear Eq.~\eqref{linear:Knn} as a function of the normalized
  wavenumber $\alpha$, such that $k=\alpha\,\pi/\Delta x$.}
  \label{fig:GroupVelocityLinear}
\end{figure}

Present results suggest that the spatial numerical approximation of
the linear term introduced in Eq.~\eqref{eq:Knn} in order to stop de
compacton may be the responsible of the self-similarity of the
envelope of the radiation studied in this paper. Introducing into
the modified equation~\cite{Villatoro99} for Method~$i$ applied to
Eq.~\eqref{eq:Knn} the solution $u=u_c+u_r$, where $u_c$ is the
compacton~\eqref{compactonsolution} and $u_r$ is the numerically
induced radiation, with $|u_r|\ll |u_c|$ in the support of the
compacton and $u_c=0$ outside it, yields
\begin{equation}
 \label{linear:Knn}
 \dpar{u_r}{t} -c_0\,
     \frac{\mathcal{B}_i(e^{\Delta x\,D})}
          {\mathcal{A}_i(e^{\Delta x\,D})}\, u_r = \mbox{h.o.t.},
\end{equation}
where $D\equiv \dparh{}{x}$ and $\mbox{h.o.t.}$ stands for
higher-order terms. The linear dispersion of this equation is
obtained by substitution of $u_r=\exp(\II\,(k\,x-w_i(k)\,t))$ into
Eq.~\eqref{linear:Knn}. The envelope of a wavepacket of radiation
propagates with the group velocity, $C_{i}(k)=\dtoth{w_i(k)}{k}$,
given by
\begin{eqnarray*}
 && C_{1}(k) = -c_0\,\dtot{}{k} \frac{\sin (\Delta x\,k)}{\Delta x}
                  = -c_0\,\cos (\Delta x\,k), \\\\
 && C_{2}(k) = -c_0\,\dtot{}{k}
    \frac{50\, \sin(\Delta x\,k) + 5\,\sin(2\,\Delta x\,k)}
         {(33+26\,\cos(\Delta x\,k)+\cos(2\,\Delta x\,k))\,\Delta x}, \\\\
 && C_{3}(k) = -c_0\,\dtot{}{k}
    \frac{100\, \sin(\Delta x\,k) + 10\,\sin(2\,\Delta x\,k)}
         {(63+56\,\cos(\Delta x\,k)+\cos(2\,\Delta x\,k))\,\Delta x}, \\\\
 && C_{4}(k) = -c_0\,\dtot{}{k}
    \frac{160\, \sin(\Delta x\,k) + 25\,\sin(2\,\Delta x\,k)}
         {(108+96\,\cos(\Delta x\,k)+6\,\cos(2\,\Delta x\,k))\,\Delta x},
\end{eqnarray*}
for Methods~1--4, respectively, which are plotted in
Figure~\ref{fig:GroupVelocityLinear} as a function of the normalized
wavenumber $\alpha$, given by $k=\alpha\,k_{\max}$ where the highest
wavenumber in the spatial grid is $k_{\max}=\pi/\Delta x$. The
discrete Fourier transform of the forward radiation shows that its
spectrum is concentrated around the highest wavenumber, therefore,
its front velocity is given by $C_{i}(k_{\max}) = c_0$, $5\,c_0$,
$10\,c_0$, and~$6.11\,c_0$, for Methods~1--4, respectively. The
spectrum for the backward radiation presents several peaks of low
frequency accompanied with a smaller peak at the highest wavenumber,
therefore, its front velocity is given approximately by
$C_{i}(k_{\max}/10) \approx -c_0$, for Methods~1--4. These results
are in good agreement with Table~\ref{VelocidadesFrentesEnCo} and
further results for the $K(p,p)$ equation omitted here for brevity.

Equation~\eqref{linear:Knn} is not the only responsible of the
numerically induced self-similar radiations by the $K(p,p)$
compactons since it is easy to show that it has no self-similar
solutions. Moreover, the numerical solution by all the methods
studied in this paper blow ups when $c_0$ either has a negative
value ($c_0\,c<0$) or has a value very different from $c$ (either
$|c_0|\ll |c|$, or $|c_0|\gg |c|$). This result, found in a large
number of simulations and illustrated in the last column of
Table~\ref{VelocidadesFrentesEnCo}, was unnoticed by the authors of
Ref.~\cite{deFrutosSanzSerna1995}, whose first introduced the linear
term in Eq.~\eqref{eq:Knn}, and in further
references~\cite{RusVillatoro2005,IsmailTaha1998,HanXu2007}.
Furthermore, a linear stability analysis of the semidiscrete
Eq.~\eqref{linear:Knn} shows its unconditional stability,
independently of the value of $c_0$, so the roots of this
instability must be in the nonlinear terms not considered in it.
%% Expresion matematica (sugerida por la var. amplitud)
%%   de solucion auto-similar

\begin{figure}
 \begin{center}
  \includegraphics[width=11.0cm]{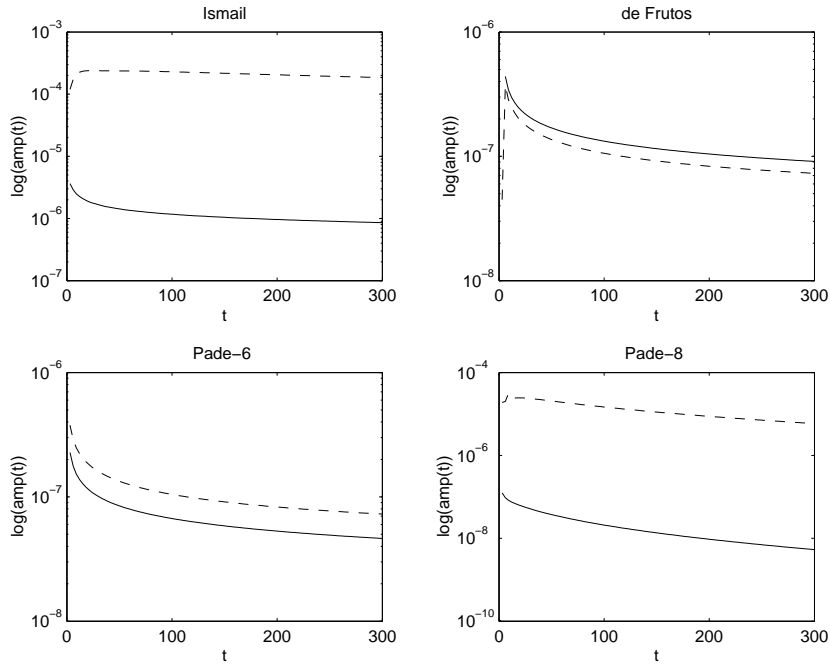} %% Amplitudes
 \end{center}
  \caption{Amplitude evolution
  of both forward (continuous line) and backward (dashed line)
  radiations as a function of time for Ismail (top left plot),
  de Frutos (top right), Pad\'e-6 (bottom left) and Pad\'e-8
 (bottom right) methods applied to the $K(2,2)$ equation.}
  \label{fig:AmplitudesRadiation}
\end{figure}

The self-similarity of both the forward and the backward radiation
requires that their analytical expressions be self-similar functions
which may be analytically written as, respectively,
\begin{eqnarray}
 \label{eq:selfsimilarForward}
 &&
   u(x,t) = t^{-\varrho_f}\,
            u_f\left( \frac{x-x_f-c\,t}{c_f\,t}\right),
   \qquad
   x_f + c\,t \le x < \infty,
 \\&&
 \label{eq:selfsimilarBackward}
   u(x,t) = t^{-\varrho_b}\,
            u_b\left( \frac{x-x_b}{|c_b|\,t}\right),
   \qquad
   \qquad
   -\infty < x \le  x_b +c\,t ,
\end{eqnarray}
where $x_b=x_0-{\pi}/({2\,\beta})$ and $x_f=x_0+{\pi}/({2\,\beta})$
are, respectively,  the left and right extremes of the compacton
solution, $\varrho_b$ and $\varrho_f$ are the scaling exponents for,
respectively, the forward and the backward radiation, and $u_f$ and
$u_b$ are the shapes of, respectively, the forward and backward
wavepackets. In order to verify that the scaling exponents in
Eqs.~\eqref{eq:selfsimilarForward}
and~\eqref{eq:selfsimilarBackward} are really constant, the temporal
evolution of the amplitude of both radiations must be studied.
Figure~\ref{fig:AmplitudesRadiation} shows that this amplitude
changes a little in the first steps of time but yields a very smooth
decreasing curve as time progresses, being approximately linear in
the logarithmic scale of the plots. Therefore, the temporal
evolution of the amplitude is asymptotically exponential in time.
Similar results have been also obtained for other $K(p,p)$
compactons and/or mesh parameters.

%% Scaling exponent => confirmando que es constante
%% Tablas factor escalado

Tables~\ref{scalingDx} and~\ref{scalingC} show the scaling exponents
in Eqs.~\eqref{eq:selfsimilarForward}
and~\eqref{eq:selfsimilarBackward} as function of $\Delta x$ and
$c_0=c$, respectively, since our extensive set of simulations show
that the time step has no significant influence on the results. The
scaling exponents in these tables have been determined by using
linear regression of the temporal evolution of the ``mean" amplitude
of the envelope of the wavepackets, i.e., the mean of the absolute
value of the amplitude of the solution in the intervals
$[x_b,x_b+c\,t]$ and $[x_f+c\,t, x_f+(c+c_f)\,t]$ for the backward
and forward radiations, respectively. To avoid the effects of the
initial transient, where the self-similarity of the wavepackets is
not properly defined due to the aliasing errors introduced by the
sampling of the solution, the first 25\% of the solution is not
considered in the linear regression.

\begin{table}
 \caption{
 Scaling exponent of both forward and backward
  radiations for the Ismail,
  de Frutos, Pad\'e-6, and Pad\'e-8
 methods for a $K(2,2)$ compacton
 in numerical simulations with $\Delta t=0.05$ and
 $c_0=c=1$ as a function of $\Delta x$, calculated using a
 linear regression of the evolution in time of the mean amplitude
 in the interval $t\in[75,300]$.
 }
{\small
\[
\begin{array}{|c|c|c|c|c|c|c|c|c|}
  \hline  %
                   & \Delta x &  0.2  &  0.1  & 0.05  & 0.025 & 0.0125 \\ \hline
  \mbox{Ismail}    &\varrho_f & 0.487 & 0.453 & 0.503 & 0.482 & 0.464 \\
                   &\varrho_b & 0.487 & 0.924 & 0.977 & 1.05 & 1.18 \\ \hline
  \mbox{de Frutos} &\varrho_f & 0.501 & 0.497 & 0.498 & 0.499 & 0.500 \\
                   &\varrho_b & 0.549 & 0.547 & 0.514 & 0.513 & 0.505 \\ \hline
  \mbox{Pad\'e-6}  &\varrho_f & 0.492 & 0.494 & 0.496 & 0.498 & 0.608 \\
                   &\varrho_b & 0.532 & 0.537 & 0.500 & 0.506 & 0.612 \\ \hline
  \mbox{Pad\'e-8}  &\varrho_f & 0.481 & 0.513 & 0.527 & 0.556 & 0.520 \\
                   &\varrho_b & 0.077 & 0.774 & 0.931 & 0.891 & 0.875 \\ \hline
 \end{array}
\]
}
 \label{scalingDx}
\end{table}

% Tabla de scaling exponent : Variando dx

Tables~\ref{scalingDx} and~\ref{scalingC} show that the scaling
factor $\varrho_f$ is approximately equal to 0.5 for all the four
methods studied in this paper, nearly independent of both $\Delta x$
and $c_0=c$, respectively, with the largest dispersion associated to
Pad\'e-8 and Ismail methods. Tables~\ref{scalingDx}
and~\ref{scalingC} also show that the scaling factor $\varrho_b$ is
approximately equal to 0.5 for the de Frutos and Pad\'e-6 methods,
to 0.9 for the Pad\'e-8 method, and to 1.0 for the Ismail method,
with a small decrement as $c$ grows, for both the Pad\'e-8 and
Ismail methods. For the last two methods the dispersion is large,
although diminish if the first column of Tables~\ref{scalingDx}
and~\ref{scalingC} is not taken into account since it corresponds to
a large $\Delta x$ and presents very noticeable aliasing errors.

\begin{table}
 \caption{
  Scaling exponent of both forward and backward
  radiations for the Ismail,
  de Frutos, Pad\'e-6, and Pad\'e-8
 methods for a $K(2,2)$ compacton
 in numerical simulations with $\Delta t=0.05$ and
 $\Delta x=0.05$ as a function of $c_0=c$,
 calculated using a linear regression of the evolution in time
 of the mean amplitude in the interval $t\in[75,300]$.
 }
{\small
\[
\begin{array}{|c|c|c|c|c|c|c|c|c|c|}
  \hline  %
                   &  c_0     &  0.1  &  0.2  &  0.5  &  1    &  1.5  &  2    &  5    \\ \hline
  \mbox{Ismail}    &\varrho_f & 0.157 & 0.512 & 0.519 & 0.503 & 0.492 & 0.487 & 0.477 \\
                   &\varrho_b & 2.408 & 1.18  & 1.07  & 0.977 & 0.827 & 0.794 & 0.778 \\ \hline
  \mbox{de Frutos} &\varrho_f & 0.492 & 0.494 & 0.496 & 0.510 & 0.495 & 0.495 & 0.498 \\
                   &\varrho_b & 0.542 & 0.552 & 0.505 & 0.511 & 0.499 & 0.495 & 0.479 \\ \hline
  \mbox{Pad\'e-6}  &\varrho_f & 0.487 & 0.489 & 0.491 & 0.496 & 0.489 & 0.490 & 0.489 \\
                   &\varrho_b & 0.561 & 0.507 & 0.437 & 0.500 & 0.496 & 0.507 & 0.504 \\ \hline
  \mbox{Pad\'e-8}  &\varrho_f & 0.534 & 0.496 & 0.562 & 0.527 & 0.530 & 0.507 & 0.482 \\
                   &\varrho_b & 0.433 & 0.962 & 0.995 & 0.931 & 0.899 & 0.760 & 0.483 \\ \hline
 \end{array}
\]
}
 \label{scalingC}
\end{table}

\section{Conclusions}
\label{Conclusions}

The propagation of compactons of the Rosenau-Hyman $K(p,p)$ equation
has been studied by means of four numerical methods showing the
appearance of numerically-induced radiation. Both backward and
forward wavepackets are generated from the compacton with a clear
self-similar shape, illustrated by means of properly scaling of the
figures. The parameters characterizing these wavepackets have been
numerically determined. An analytical model, based on the
linearization of $K(p,p)$ equation has been used in order to
approximate the front velocities of the forward wavepackets relative
to that of the compacton. The front velocity of the backward
wavepacket is approximately equal to that of the compacton but with
opposite sign. Both forward and backward front velocities are
practically independent of the parameters $\Delta x$ and $\Delta t$
of the numerical method. The evolution in time of the mean amplitude
of the wavepackets shows its exponential decreasing in time,
suggesting an ansatz for both the backward and forward wavepackets
corresponding to self-similar functions characterized by scaling
exponents, which have been numerically calculated for both
radiations by means of the linear regression of the logarithm of the
mean amplitude, showing that its value is approximately constant as
a function of both the mesh grid size and the compacton velocity.

The scaling exponents approximately equal to $1/2$ for the amplitude
evolution of the envelope of both the forward and the backward
radiations suggest that they may be analytically approximated for
weak nonlinearity (valid for the radiation but not for the
compactons) by a nonlinear (cubic) Schr\"odinger equation. This
analysis is in progress. Asymptotic analysis using the method of
modified equations applied to the four numerical methods studied in
this paper may also be useful in the characterization of the
radiation wavepackets found here. In fact, the analytical
explanation of both the generation of the self-similar radiations
and the blow-up of the solution for some $c_0$ are very interesting
open problems.

%%%%%%%%%%%%%%%%%%%%%%%%%%%%%%%%%%%%%%%%%%%%%%%%%%%%
%%%%%%%%%%%%%%%%%%%%%%%%%%%%%%%%%%%%%%%%%%%%%%%%%%%%
\section*{Acknowledgments}

The authors thanks the referees of this paper for its useful remarks
which have greatly improved the paper. The research reported in this
paper was partially supported by Projects FIS2005-03191 and
TIN2005-09405-C02-01 from the Ministerio de Educaci\'on y Ciencia,
Spain.

%% TIN2005-09405-C02-01 DESARROLLO DE SOFTWARE PARA SISTEMAS
%% DISTRIBUIDOS P2P 2 MANUEL DIAZ RODRIGUEZ mdr@lcc.uma.es 952131394
%% UNIVERSIDAD DE MALAGA ESCUELA TECNICA SUPERIOR
%% DE INGENIERIA INFORMATICA Málaga

%%%%%%%%%%%%%%%%%%%%%%%%%%%%%%%%%%%%%%%%%%%%%%%
%%%%%%%%%%%%%%%%%%%%%%%%%%%%%%%%%%%%%%%%%%%%%%%
%PUT HERE THE JOURNAL BIBLIOGRAPHY CONVENTIONS.
% Without using the BIBITEX facility.

%% References JOURNAL OF COMPUTATIONAL PHYSICS (everything without italics)
%% References should be denoted in the text by numbers in brackets and listed, with full paper titles, at the end of the paper in numerical order, in the following style:
%% 1. J. Szumbarski and J. M. Floryan, A direct spectral method for determination of flows over corrugated boundaries, J. Comput. Phys. 153, 378 (1999), doi:10.1006/jcph.1999.6282.
%% 2. P. Betsch and P. Steinmann, Inherently energy conserving time finite elements for classical mechanics, J. Comput. Phys. (2000), doi:10.1006/jcph.1999.6427.
%% 3. I. Gohberg, P. Lancaster, and L. Rodman, Matrix Polynomials (Academic Press, New York, 1982), n. 54.
%% 4. E. F. Toro, Riemann-problem based techniques for computing reactive two-phase flows, in Proc. Third Intern. Confer. on Numerical Combustion, Antibes, France, May 1989, edited by Dervieux and Larrouturrou, Lect. Notes in Phys. (Springer- Verlag, New York/Berlin, 1989), Vol. 351, p. 472

%\bookref{authors,title,editorial,year,city,pp}
%\bookref{   1   ,  2  ,    3    , 4  ,  5 , 6}
%% [2] Kwang K and Briggs FA. Computer architecture and parallel processing.
%%        New York: McGraw-Hill, 1984, pp. 10-23.
\newcommand{\bookref}[5]{ {#1}, {#2} ({#3}, {#5}, {#4}).}
\newcommand{\bookrefpag}[6]{ {#1}, {#2} ({#3}, {#5}, {#4}), p.~{#6}.}

%\paperref{authors,title,journal,vol,numb,pp,year,shortname}
%             1      2      3     4    5   6   7      8
% sin puntos autores y nombre revista completo
%% [1] McBryan O. Matrix and vector operations in hypercube parallel processors.
%%        Parallel Computing, 1987;5:117-125.
\newcommand{\paperref}[8]{ {#1}, {#2}, {#8}, {#4} ({#7}) {#6}.}
\newcommand{\paperrefno}[8]{ {#1}, {#2}, {#8}, {#4}(#5) ({#7}) {#6}.}
\newcommand{\paperrefart}[8]{ {#1}, {#2}, {#8}, {#4} (#7) Art.~{#6}.}


\begin{thebibliography}{00}

 \bibitem{RosenauHyman1993} \paperrefno
 {P. Rosenau and J. M. Hyman}
 {Compactons: Solitons with finite wavelength}
 {Physical Review Letters}
 {70} {5} {564--567} {1993}{Phys. Rev. Lett.}

 \bibitem{Rosenau1998} \paperrefno
 {P. Rosenau}
 {On a class of nonlinear dispersive-dissipative
  interactions}
 {Physica D}
 {123} {1--4} {525--546} {1998} {Physica D}

 \bibitem{Ludu1998} \paperrefno
 {A. Ludu and J. P. Draayer}
 {Patterns on liquid surfaces: cnoidal waves, compactons and scaling}
 {Physica D}
 {123} {} {82--91} {1998}{Physica D}

 \bibitem{Bertozzi1996} \paperrefno
 {A. L. Bertozzi and M. Pugh}
 {The lubrication approximation for thin viscous films: regularity and
  long time behavior of weak solutions}
 {Communications on Pure and Applied Mathematics}
 {49} {2} {85--123} {1996}{Commun. Pure Appl. Math.}

 \bibitem{Kovalev1998} \paperrefno
 {A. S. Kovalev and M. V. Gvozdikova}
 {Bose gas with nontrivial particle interaction and semiclassical
  interpretation of exotic solitons}
 {Low Temperature Physics}
 {24} {7} {484--488} {1998}{Low Temp. Phys.}

 \bibitem{Grimshaw1998} \paperrefno
 {R. H. J. Grimshaw, L. A. Ostrovsky, V. I. Shrira, and Y. A. Stepanyants}
 {Long nonlinear surface and internal gravity waves in a rotating ocean}
 {Surveys in Geophysics}
 {19} {4} {289--338} {1998}{Surv. Geophys.}

 \bibitem{Kardashov2006} \paperrefart
 {V. Kardashov, S. Einav, Y. Okrent, and T. Kardashov}
 {Nonlinear reaction-diffusion models of self-organization and
 deterministic chaos: Theory and possible applications to description
 of electrical cardiac activity and cardiovascular circulation}
 {Discrete Dynamics in Nature and Society}
 {2006} {} {98959} {2006}{Discrete Dyn. Nat. Soc.}

 \bibitem{Simpson2007} \paperref
 {G. Simpson, M. Spiegelman, and M. I. Weinstein}
 {Degenerate dispersive equations arising in the study of magma dynamics}
 {Nonlinearity}
 {20} {} {21--49} {2007}{Nonlinearity}

 \bibitem{Caparelli1998} \paperref
 {E. C. Caparelli, V. V. Dodonov, and S. S. Mizrahi}
 {Finite-length soliton solutions of the local homogeneous nonlinear
 Schrödinger equation}
 {Physica Scripta}
 {58} {} {417--420} {1998}{Phys. Scr.}

 \bibitem{Dusuel1998} \paperrefno
 {S. Dusuel, P. Michaux, and M. Remoissenet}
 {From kinks to compactonlike kinks}
 {Physical Review E}
 {57} {2} {2320--2326} {1998}{Phys. Rev. E}

 \bibitem{Comte2002} \paperref
 {J. C. Comte}
 {Compact traveling kinks and pulses}
 {Chaos, Solitons \& Fractals}
 {14} {} {1193--1199} {2002}{Chaos Solitons Fractals}

 \bibitem{Comte2006} \paperref
 {J. C. Comte and P. Marqui\'e}
 {Compact-like kink in real electrical reaction-diffusion chain}
 {Chaos, Solitons \& Fractals}
 {29} {} {307--312} {2006}{Chaos Solitons Fractals}

 \bibitem{Prilepsky2006} \paperrefart
 {J. E. Prilepsky, A. S. Kovalev, M. Johansson, and Y. S. Kivshar}
 {Magnetic polarons in one-dimensional antiferromagnetic chains}
 {Physical Review B}
 {74}{} {132404} {2006}{Phys. Rev. B}

 \bibitem{Pikovsky2005} \paperrefart
 {P. Rosenau and A. Pikovsky}
 {Phase compactons in chains of dispersively coupled oscillators}
 {Physical Review Letters}
 {94} {} {174102} {2005}{Phys. Rev. Lett.}

 \bibitem{Pikovsky2006} \paperref
 {A. Pikovsky and P. Rosenau}
 {Phase compactons}
 {Physica D}
 {218} {} {56--69} {2006}{Physica D}

 \bibitem{Rosenau2000} \paperrefno
 {P. Rosenau}
 {Compact and noncompact dispersive patterns}
 {Physics Letters A}
 {275} {3} {193--203} {2000} {Phys. Lett. A}

 \bibitem{Cooper2006} \paperrefno
 {F. Cooper, A. Khare, and A. Saxena}
 {Exact elliptic compactons in generalized Korteweg–De Vries equations}
 {Complexity}
 {11} {6}{30--34} {2006}{Complexity}

 \bibitem{Rosenau2006} \paperrefno
 {P. Rosenau}
 {On a model equation of traveling and stationary compactons}
 {Physics Letters A}
 {356} {1} {44--50} {2006} {Phys. Lett. A}

 \bibitem{RosenauHyman2007} \paperrefart
 {P. Rosenau, J. M. Hyman, and M. Staley}
 {Multidimensional compactons}
 {Physical Review Letters}
 {98} {} {024101} {2007}{Phys. Rev. Lett.}

 \bibitem{deFrutosSanzSerna1995} \paperrefno
 {J. de Frutos, M. A. L\'opez-Marcos, and J. M. Sanz-Serna}
 {A finite difference scheme for the $K(2,2)$ compacton equation}
 {Journal of Computational Physics}
 {120} {2} {248--252} {1995} {J. Comput. Phys.}

 \bibitem{SaucezEtAl2004} \paperrefno
 {P. Saucez, A. V. Wouwer, W. E. Schiesser, and P. Zegeling}
 {Method of lines study of nonlinear dispersive waves}
 {Journal of Computational and Applied Mathematics}
 {168} {1-2} {413--423} {2004} {J. Comput. Appl. Math.}

 \bibitem{ChertockLevy2001} \paperrefno
 {A. Chertock and D. Levy}
 {Particle methods for dispersive equations}
 {Journal of Computational Physics}
 {171} {2} {708--730} {2001}{J. Comput. Phys.}

 \bibitem{CooperHyman2001} \paperrefart
 {F. Cooper, J. M. Hyman, and A. Khare}
 {Compacton solutions in a class of generalized fifth-order
  Korteweg-de Vries equations}
 {Physical Review E}
 {64} {2} {026608} {2001} {Phys. Rev. E}

 \bibitem{RusVillatoro2005}
 {F. Rus and F. R. Villatoro},
 {Pad\'e Numerical method for the Rosenau-Hyman compacton
 equation}, {Math. Comput. Simul.} (2007), doi:10.1016/ j.matcom.2007.01.016.

 \bibitem{LevyShuYan2004} \paperrefno
 {D. Levy, C. W. Shu, and J. Yan}
 {Local discontinuous Galerkin methods for nonlinear dispersive equations}
 {Journal of Computational Physics}
 {196} {2} {751--772} {2004}{J. Comput. Phys.}

 \bibitem{IsmailTaha1998} \paperrefno
 {M. S. Ismail and T. R. Taha}
 {A numerical study of compactons}
 {Mathematics and Computers in Simulation}
 {47} {6} {519--530} {1998} {Math. Comput. Simul.}

 \bibitem{HanXu2007} \paperrefno
 {H. Han and Z. Xu}
 {Numerical solitons of generalized Korteweg-de Vries equations}
 {Applied Mathematics and Computation}
 {186} {1} {483--489} {2007} {Appl. Math. Comput.}


 \bibitem{RusVillatoro2006} \paperrefno
 {J. Garral\'on, F. Rus, and F. R. Villatoro}
 {Compacton numerically-induced radiation in a fourth-order finite element method}
 {WSEAS Transactions on Mathematics}
 {5} {1} {89--96} {2006} {WSEAS T. Math.}


 \bibitem{Villatoro99} \paperrefno
 {F. R. Villatoro and J. I. Ramos}
 {On the method of modified equations. I: Asymptotic analysis of the
  Euler forward difference method}
 {Applied Mathematics and Computation}
 {103} {2-3} {111--139} {1999} {Appl. Math. Comput.}

\end{thebibliography}
\end{document}